\documentclass{article}

\usepackage{arxiv}

\usepackage[utf8]{inputenc} 
\usepackage[T1]{fontenc}    
\usepackage{hyperref}       
\usepackage{url}            
\usepackage{booktabs}       
\usepackage{amsfonts,amsmath}       
\usepackage{nicefrac}       
\usepackage{microtype}      
\usepackage{lipsum}

\usepackage{graphicx}
\usepackage{cuted,multirow,comment,cite}
\usepackage{adjustbox}
\usepackage{mdframed,fancyvrb}
\usepackage{listings}
\usepackage{multicol,ulem}
\usepackage{subcaption}
\definecolor{codegreen}{rgb}{0,0.6,0}
\definecolor{codegray}{rgb}{0.5,0.5,0.5}
\definecolor{codepurple}{rgb}{0.58,0,0.82}
\definecolor{backcolour}{rgb}{0.95,0.95,0.92}
\lstdefinestyle{mystyle}{
    backgroundcolor=\color{backcolour},   
    commentstyle=\color{codegreen},
    keywordstyle=\color{magenta},
    numberstyle=\tiny\color{codegray},
    stringstyle=\color{codepurple},
    basicstyle=\ttfamily\scriptsize,
    breakatwhitespace=false,         
    breaklines=true,                 
    captionpos=b,                    
    keepspaces=true,                 
    numbers=left,                    
    numbersep=5pt,                  
    showspaces=false,                
    showstringspaces=false,
    showtabs=false,                  
    tabsize=2
}
\title{Computer based activity to understand \\ proper acceleration using the Rindler observer}

\author{
  Sushil~Kumar~Singh \\
  Department of Physics \\SGTB Khalsa College\\
  University of Delhi
   \And
 Gervit~Kumar~Trehan \\
  Department of Physics\\ SGTB Khalsa College\\
  University of Delhi
   \And
 Savinder ~Kaur\thanks{sk\_savinder2005@yahoo.co.in} \\
  Department of Physics\\ SGTB Khalsa College\\
  University of Delhi
}

\begin{document}
\maketitle

\begin{abstract}
Using elementary knowledge of Special Relativity, we design a computational classroom experiment in excel and python. Here, we show that any inertial observer $\mathcal{B}$ with an arbitrary speed $u_\mathcal{B}$ is associated with a unique event ${E}$ along the worldline of Rindler observer $\mathcal{R}$. The inertial observer $\mathcal{B}$ records the velocity of $\mathcal{R}$ to be non-zero at every instant except at this unique ${E}$ where it is co-moving with $\mathcal{R}$. At this event, $\mathcal{B}$ records a minimized distance to $\mathcal{R}$ and a maximized acceleration of $\mathcal{R}$ along its worldline. Students grasp the concept of proper acceleration when they realise that though an inertial observer measures variable local acceleration but this maxima is the same for all inertial observers. Since the Rindler observer is associated with variable local velocity the time dilation factors are different. Parameterising the Rindler velocity by proper time we graphically present the concept of time dilation. We assume $\mathcal{B}$ to be moving with respect to an inertial observer $\mathcal{A}$ which is at rest and their clocks synchronize to zero when they meet.

\end{abstract}

\keywords{ Rindler Observer \and Proper acceleration \and Comoving inertial frames \and Time dilation \and Computer based experiment}

\section{Introduction}
Mermin \cite{Mermin-1997} uses geometrical approach exploiting symmetries to introduce Minkowski spacetime while Salgado \cite{Salgado-2016} constructs spacetime diagrams for non-scientists using a graph paper. Misner Thorne and Wheeler \cite {Misner-2017}, Rindler \cite{Rindler-1977} and Mukhanov \cite{Mukhanov-2007} present an algebraic approach to the case of rectilinear motion with constant proper acceleration (i.e. to explain a Rindler observer $\mathcal{R}$). 
Flores\cite{Flores-2007} discusses communication amongst inertial and accelerated observers for beginners. Hughes\cite{Hughes-2021} in a recent paper discusses time dilation by presenting a thought experiment. Semay\cite{Semay-2006} presents several aspects of spacetime of the uniformly accelerated observer without resorting to general relativity. 

Non availability of class room activities dealing with concepts of instantaneously comoving inertial frames and proper acceleration has propelled us to design a hands-on experiment. We use the notion of proper time, to elucidate the measurement of time intervals by inertial and accelerated observers and extend our understanding of time dilation. Employing elementary algebra we develop simple codes in excel as well as in python that allows the undergraduates to explore such ideas and enhance their understanding for $\mathcal{R}$. We assume students are familiar with spacetime diagrams\cite{Takeuchi-2010,Bais-2007,Hassani-2017} and the Lorentz transformation\cite{Hassani-2017}.

We organise this paper with Section \ref{sec2} discussing  a worldline of $\mathcal{R}$ with constant spacetime interval from an inertial observer $\mathcal{A}$ at rest and derive its coordinate velocity and acceleration. In Section \ref{sec3} we show that for every event along the worldline of $\mathcal{R}$ there is a unique inertial observer $\mathcal{B}$ moving with speed $u$ relative to $\mathcal{A}$ and confirms that $\mathcal{B}$ is comoving with $\mathcal{R}$ at that time slice. In Section \ref{sec4} we present the quantitative results alongside the computational codes of the hands-on experiment. The codes use (i) the Lorentz transformation (ii) yield the coordinate velocity and acceleration using finite difference (iii) the separation of time between two fixed events measured by $\mathcal{B}$ and $\mathcal{R}$ and (iv) at a `conceptual' level interpret proper acceleration and time dilation. We conclude in Section \ref{sec5}.

\section{Rindler Observer \& Constant Spacetime Interval}\label{sec2}
Let an inertial observer $\mathcal{A}$ at rest register a series of discrete events $E_n$ with $n=0, 1, 2, 3 \dots$ happening at times $t_n$ at different places $x_{n}$ from itself
\begin{align} \label{s2e1}
x\left( {E_n} \right) = x_{n}=\sqrt{c^2t_n^2+l_0^2}
\end{align} 
here, $l_0$ is the distance of nearest event $E_0~ (t_0 = 0)$. The spacetime interval $s$ from the present event $O$ ( $t_n=0$) to all these events $E_n$ are
\begin{align} \label{s2e2}
s^2\left( {E_n} \right) = x_n^2 - c^2 t_n^2 = l_0^2
\end{align}
where $\left(x_n,ct_n\right)$ are the spacetime coordinates of $E_n$. 
\begin{figure}[!ht]
\centering
    \includegraphics[width=0.47\textwidth]{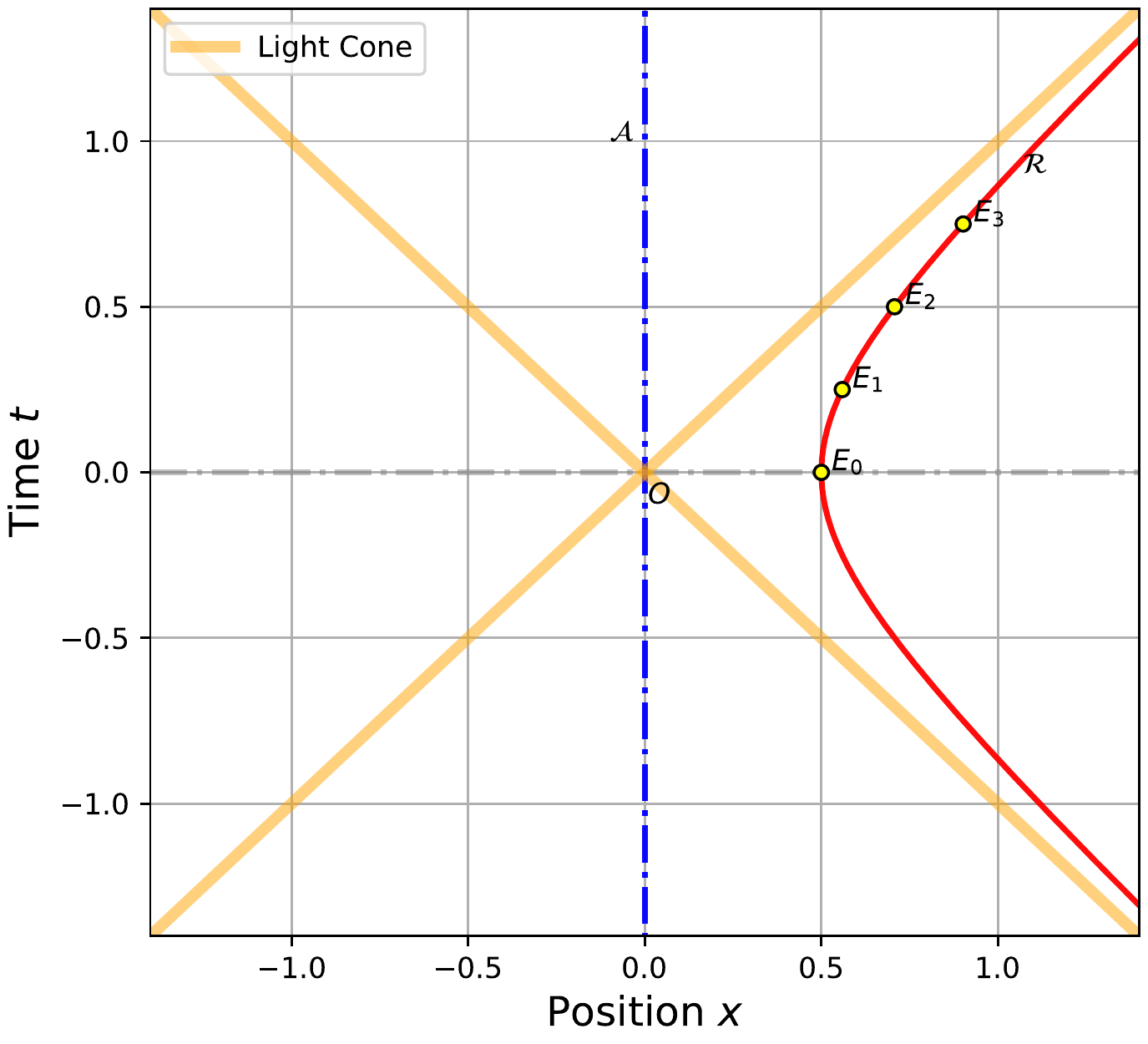}
\caption{Spacetime of $\mathcal{A}$ and worldline of $\mathcal{R}$.}
\label{fig:1}
\end{figure}
The event $E_{n+1}$ being timelike to event $E_n$ i.e.
\begin{align*}
  x_{n+1}^2-x_{n}^2 &= c^2(t_{n+1}^2-t_{n}^2) \rightarrow x_{n+1}-x_{n} = c(t_{n+1}-t_{n})\left(\frac{ct_{n+1}+ct_{n}}{x_{n+1}+x_{n}}\right)  \\
 &\xrightarrow[]{x_n< ct_n,~x_{n+1}< ct_{n+1}} ~ x_{n+1}-x_{n} < c~(t_{n+1}-t_{n})
\end{align*}
another observer $\mathcal{R}$ (called the Rindler observer) moves in such a manner that he witnesses these event as he passes through each one of them. Figure \ref{fig:1} shows the spacetime of $\mathcal{A}$ with $t_n=n\Delta t$ for $\Delta t=0.25$. The worldlines of photons transgressing are shown in yellow (solid) and that of inertial observer A in blue (dash dot). In the space-time of $\mathcal{A}$, the world-line of $\mathcal{R}$ will be a hyperbola (red curve in Fig \ref{fig:1})
\begin{equation}\label{s2e3} 
X^2 - c^2 T^2 = l_0^2
\end{equation}
where $\left(X,cT\right)$ are the spacetime coordinates of all events along the worldline of $\mathcal{R}$. By taking the time derivative of Eq.\eqref{s2e3} we write
\begin{equation}\label{s2e4}
{X}~ \frac{dX}{dT} - c^2 T = 0
\end{equation}
the speed of $\mathcal{R}$ as measured by the inertial observer $\mathcal{A}$ is
\begin{equation}\label{s2e5} 
\frac{dX}{dT} = \frac{c^2T}{X} = \pm ~c\sqrt{ 1-\frac{l_0^2}{X^2} }
\end{equation}
and taking the time derivative of Eq.\eqref{s2e5} the acceleration is
\begin{equation}\label{s2e6} 
\frac{d^2X}{ dT^2 } = \frac{c^2 l_0^2}{X^3}  
\end{equation}
As $\mathcal{R}$ moves away from $\mathcal{A}$ its speed increases but with diminishing acceleration. The speed of $\mathcal{R}$ however, is bound by the speed of light 
\begin{align}
\frac{dX}{dT} = c\sqrt{ 1-\frac{l_0^2}{X^2} }~~ \xrightarrow[]{X\rightarrow\infty} c
\end{align}

\section{Instantaneously Comoving Inertial Frames}\label{sec3}

The spacetime interval being an invariant quantity, interval $s^{\prime}$ of the discrete events $E_n$ remains the same as $s$ for any other inertial observer $\mathcal{B}$ which is moving with an arbitrary speed $u$ with respect to $\mathcal{A}$. 
\begin{align} \label{s3e1}
s^{\prime~2} = x_n^{\prime~2} - c^2 t_n^{\prime~2} = l_0^2
\end{align}
The Lorentz transformations, in general
\begin{align} 
x = \gamma_u \left( x^\prime + \beta_u ct^\prime \right) \label{s3e2}\\
ct = \gamma_u \left( ct^\prime + \beta_u x^\prime \right) \label{s3e3}
\end{align}
provide the relation between the coordinates $(x^\prime, ct^\prime)$ and $(x, ct)$ of any event $E$ as observed by the moving observer $\mathcal{B}$ and the observer $\mathcal{A}$ respectively. Here $\beta_u=u/c$ and $\gamma_u= 1/\sqrt{1-\beta_u^2}$. 

We show that an inertial observer $\mathcal{B}$ with speed $u$ is uniquely associated with an event along the worldline of the Rindler observer. Equivalently, for any discrete event $E_n$ along the worldine of $\mathcal{R}$ there is always a unique inertial observer $\mathcal{B}_n$ with speed $u_n$ which is momentarily comoving with the $\mathcal{R}$. 
\begin{enumerate}
\item We can make the event $O$ simultaneous to the event $E_n$ (i.e. the time coordinate $t^{\prime}_n=0$) for the inertial observer $\mathcal{B}_n$ with an appropriate speed $u_n$. The space coordinate of $E_n$ from  Eq.\ref{s3e1} is then
\begin{equation}\label{s3e4} 
x^\prime_n = l_0 
\end{equation}
assuming the events to be happening at the positive space coordinates. Using these spacetime coordinates, the Lorentz transformations Eq.\eqref{s3e2} and Eq.\eqref{s3e3} give relations $x_n = \gamma_u \left( x_n^{\prime} \right) = \gamma_u l_0$ and $ct_n=\gamma_u \left( \beta_u x_n^{\prime} \right) = \gamma_u \beta_u l_0$ from which we get 
\begin{align} \label{s3e5}
\beta_u = \frac{ct_n}{x_n} = \frac{c t_n}{ \sqrt{l_0^2+c^2t_n^2 }} \rightarrow u_n =  c\frac{c t_n}{ \sqrt{l_0^2+c^2t_n^2 }}
\end{align}
as the speed of the inertial observer $\mathcal{B}_n$. 
\item  This also makes the $\mathcal{B}_n$ share a unique relation with $\mathcal{R}$. From Eq.\ref{s2e5}, the instantaneous speed of $\mathcal{R}$ as measured by $\mathcal{B}_n$ at the event $E_n$ is  
\begin{equation}\label{s3e6} 
\frac{dX}{dT}\Bigg|_{E_n} = \frac{c^2 T}{X}\Bigg|_{E_n} = \frac{c^2 t^\prime_n}{x^\prime_n} = 0
\end{equation}
The observer $\mathcal{B}_n$ is momentarily ``comoving'' with $\mathcal{R}$.
\item In addition, the instantaneous ``proper'' acceleration from Eq.\ref{s2e6} is 
\begin{equation}\label{s3e7} 
\frac{d^2X}{d{T}^2}\Bigg|_{E_n} = \frac{c^2 l_0^2}{X^3} \Bigg|_{E_n} = \frac{c^2 l_0^2}{x^{\prime 3}_n} = \frac{c^2}{l_0}
\end{equation}

For an arbitrary discrete event $E_n$ and consequently the observer $B_n$, the instantaneous proper acceleration measured is the same for all $B_n$.  Since the $B_n$ are associated with unique but different events $E_n$ along the worldline of $\mathcal{R}$, the acceleration measured by them is the `proper' acceleration felt by $\mathcal{R}$ all along the worldline.  
\end{enumerate}

\section{Computation Results}\label{sec4}

We use spacetime coordinates $(X,cT)$ of the Rindler observer relative to $\mathcal{A}$ and the Lorentz Transformation to determine the spacetime coordinates $(X^\prime,cT^\prime)$ of $\mathcal{R}$ relative to the inertial observers $\mathcal{B}$. From Eq.\eqref{s3e5} we determine the velocities $u=c\beta$ of the inertial observer $\mathcal{B}$ and draw the worldline and the space-axis of $\mathcal{B}$ (Fig.\ref{fig:2a}). We determine the velocity $V^{\prime}$ of the Rindler observer as observed by $\mathcal{B}$ by numerically differentiating the position $X^{\prime}$ with respect to $T^{\prime}$. For the acceleration $A^{\prime}$ we numerically differentiate velocity $V^{\prime}$ with respect to $T^{\prime}$. For the purpose of illustration, in our calculations, we take $c=1$, the spacetime interval $l_0=0.5$. For Fig \ref{fig:1} the following two code snippets are used to draw the spacetime of $\mathcal{A}$.

\begin{lstlisting}[language = Python]
# For worldline of A
# 't' is a user-defined list containing the values of time recorded by A
def xO_O(t): 
    xO=[]
    for i in t:
        xO.append(0)
    return xO
# For worldline of R and events En as measured by A
# 's' is the spacetime interval to be defined by the user 
def xR_O(t,s):
    xR=[]
    for i in t:
        x=np.sqrt(s**2+i**2); xR.append(x);
    return xR
\end{lstlisting}


\subsection{Proper Acceleration}\label{s4sub1}

\begin{figure}[!ht]
\centering
\begin{subfigure}{0.49\textwidth}
    \includegraphics[width=0.95\textwidth]{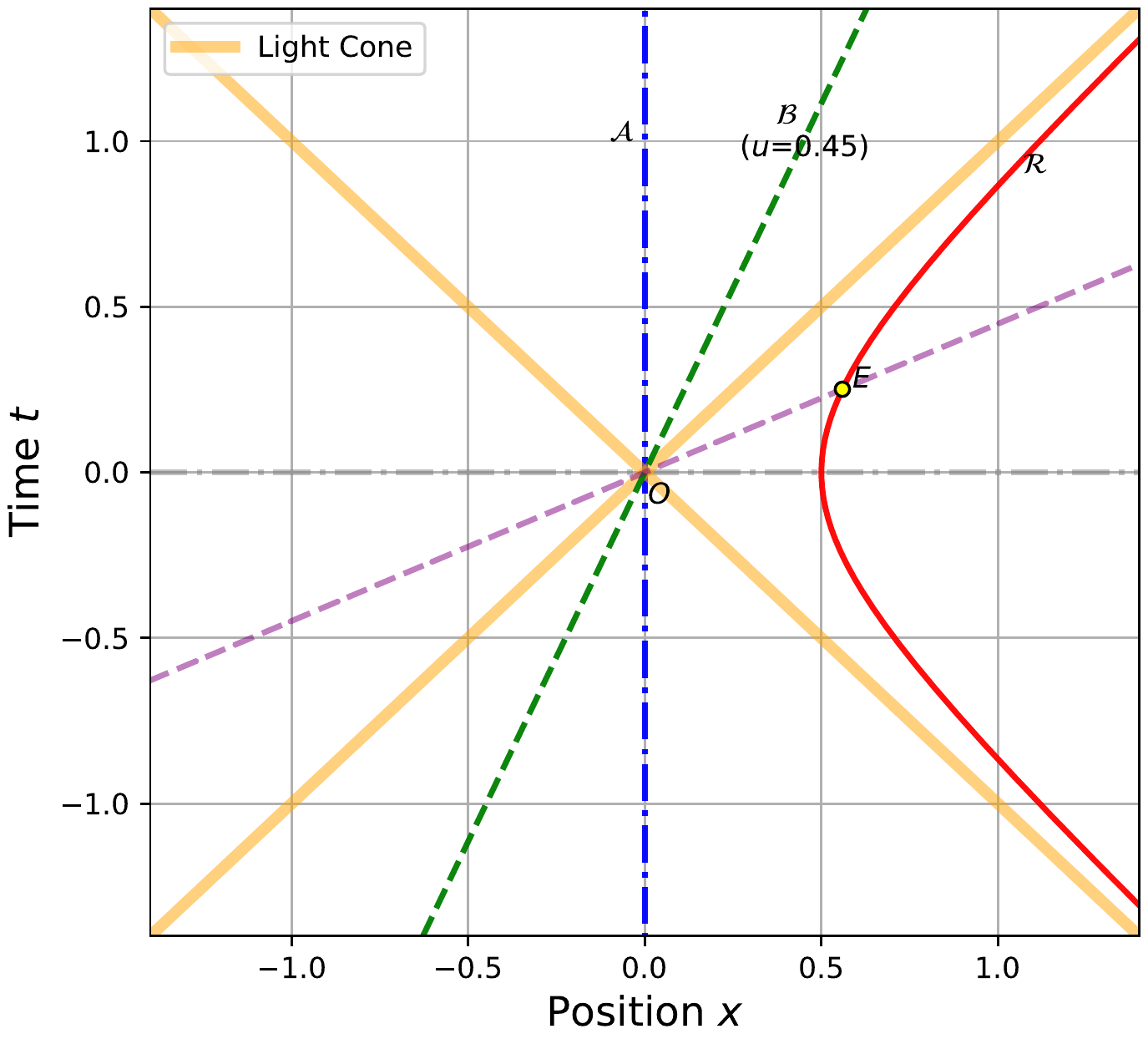}
    \caption{\scriptsize{Worldlines of $\mathcal{A}$, $\mathcal{B}$, $\mathcal{R}$ and the lightcone in the Spacetime of $\mathcal{A}$.}}
    \label{fig:2a}
\end{subfigure}
\hfill
\begin{subfigure}{0.49\textwidth}
    \includegraphics[width=0.95\textwidth]{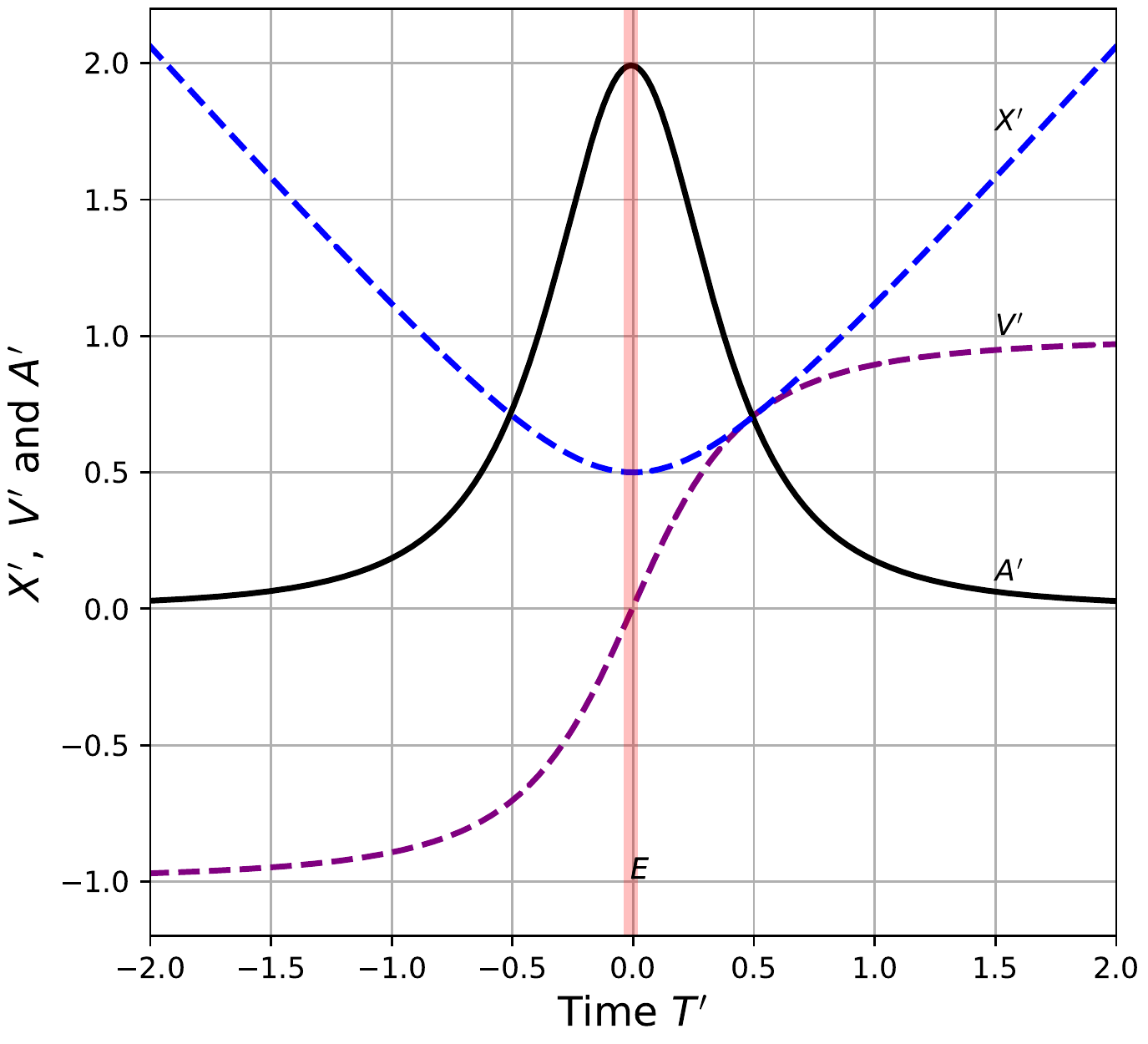}
    \caption{\scriptsize{Distance $X^{\prime}$, speed $V^{\prime}$ and acceleration $A^{\prime}$ of $\mathcal{R}$ as measured by $\mathcal{B}$}}
    \label{fig:2b}
\end{subfigure}
\caption{Spacetime of $\mathcal{A}$ and observation made by $\mathcal{B}$}
\label{fig:2}
\end{figure}
Figure \ref{fig:2a} shows the spacetime of observer $\mathcal{A}$ with worldlines of $\mathcal{A}$ at rest and $\mathcal{B}$ with arbitrary speed $u_{\mathcal{B}}=0.45$ and $\mathcal{R}$. For brevity, we shall associate the speed as $u_\mathcal{B}$ for any inertial observer $\mathcal{B}$. For $\mathcal{B}$, the event $E$ on worldline of $\mathcal{R}$ is simultaneous with event $O$ i.e. $T^{\prime}(E)=0$. Figure \ref{fig:2b} shows that the relative speed of Rindler observer $V^{\prime}$ with respect to $\mathcal{B}$ is $zero$ at the event $E$. At this event the acceleration $A^{\prime}$ of $\mathcal{R}$ measured by the observer $\mathcal{B}$ is $2.0$ as exhibited by the maximum in the curve. This value is inverse of the assumed spacetime interval $l_0=0.5$. Also at this event, $\mathcal{R}$ is closest to  $\mathcal{B}$ where $X^{\prime}(E)=l_0= 0.5$. The observer $\mathcal{B}$ is co-moving with $\mathcal{R}$ only at this event $E$ since $V^{\prime}(E)=0$ and the corresponding ``proper'' acceleration $A^{\prime}(E)$ is $2.0$. 

Figure \ref{fig:2b} although plotted for $u_B=0.45$ remains unchanged for any other speed $|u_B| < 1$ and pertains to any observer $\mathcal{B}$. We see that all inertial observers $\mathcal{B}$ are equivalent and the Rindler observer appears the same with it coming closest to $0.5$ at time $T^{\prime}=0$  where it is comoving and having proper acceleration $2$. In addition to code used for Fig \ref{fig:1} the following code snippets are used in Fig \ref{fig:2a} and Fig \ref{fig:2b}.
\begin{lstlisting}[language = Python]
# to calculate the velocity and acceleration
# x represents the independent variable (time T or T') and y represents the dependent variable (position or velocity)
def deriv(x,y): 
    rate=[]
    for i in range(len(x)-1):
        dx=x[i+1]-x[i]; dy=y[i+1]-y[i]; rate.append(dy/dx);
    return rate
# Lorentz transformation of coordinates of events Ei from A to B
# Speed u_I of the inertial observer B is obtained using Eq. 11, t and x are lists for the corresponding time and position coordinates as recorded by A
def lorentz(u_I,t,x): 
    gamma=1/np.sqrt(1-u_I**2); #print(gamma)
    xRI=[];tRI=[];
    for i in range(0,len(t)):
        x1=x[i]-u_I*t[i]; x1=gamma*x1;
        t1=t[i]-u_I*x[i]; t1=gamma*t1;
        xRI.append(x1); tRI.append(t1);
    return xRI,tRI

def IandR(s,u_I,t):    
    xIO=u_I*t; # worldline of B observed by A
    spaceIO=(1/u_I)*t; # space axis of B in the A spacetime
    xRO=xR_O(t,s); 
    vRO=t/xRO; # velocity of R measured by A
    rap=np.arctanh(vRO); # rapidity for R
    return xIO,spaceIO,xRO,vRO,rap
\end{lstlisting}

\subsection{Proper Time and Time Dilation}\label{s4sub2}
\begin{figure}[!ht]
\centering
\begin{subfigure}{0.49\textwidth}
    \includegraphics[width=0.95\textwidth]{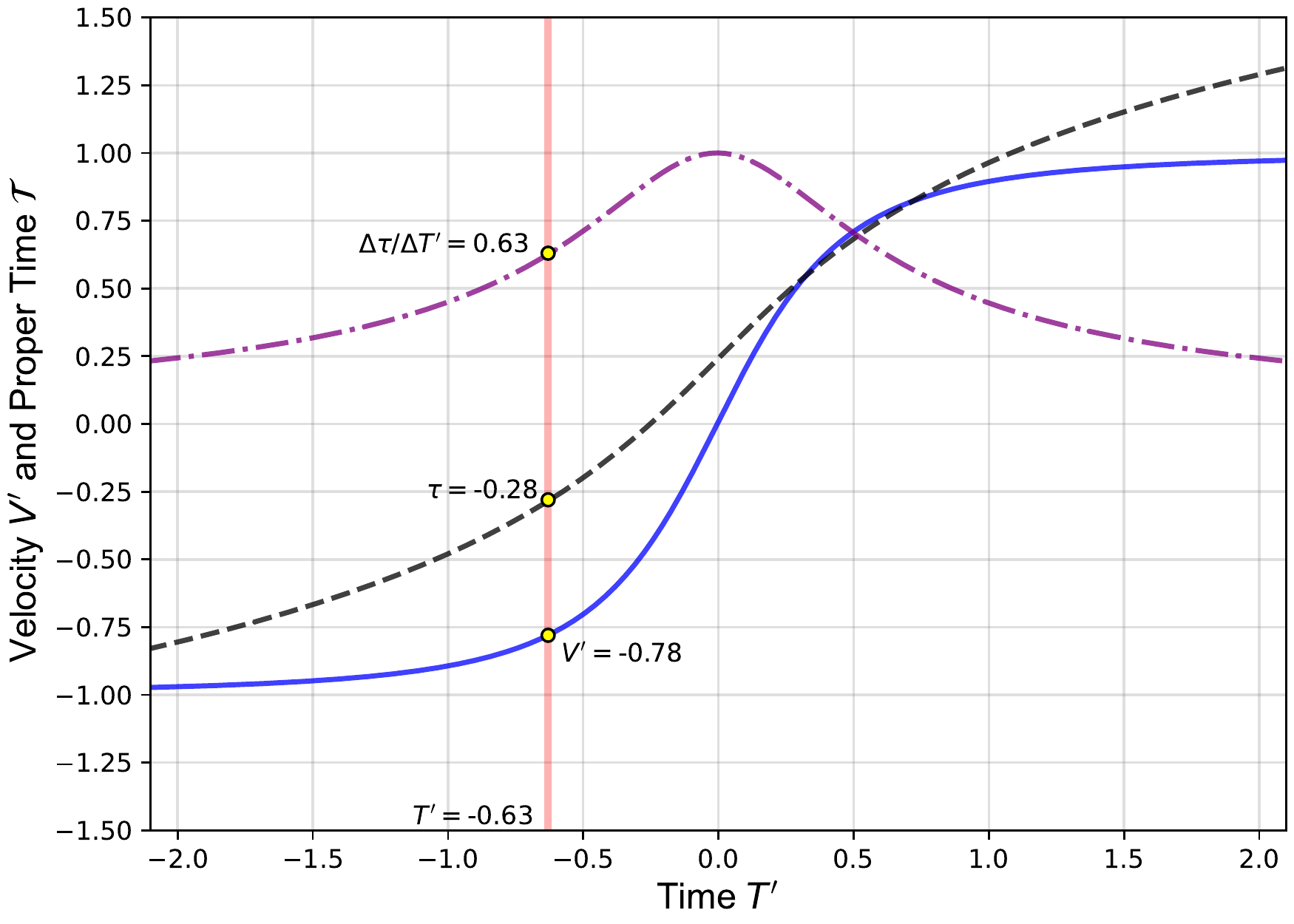}
    \caption{\scriptsize{Velocity $V^\prime$ and Proper time $\tau$ with time $T^\prime$ as measured by any inertial observer $\mathcal{B}$ with speed $u_\mathcal{B}=0.45$.} }
    \label{fig:3a}
\end{subfigure}
\hfill
\begin{subfigure}{0.49\textwidth}
    \includegraphics[width=0.95\textwidth]{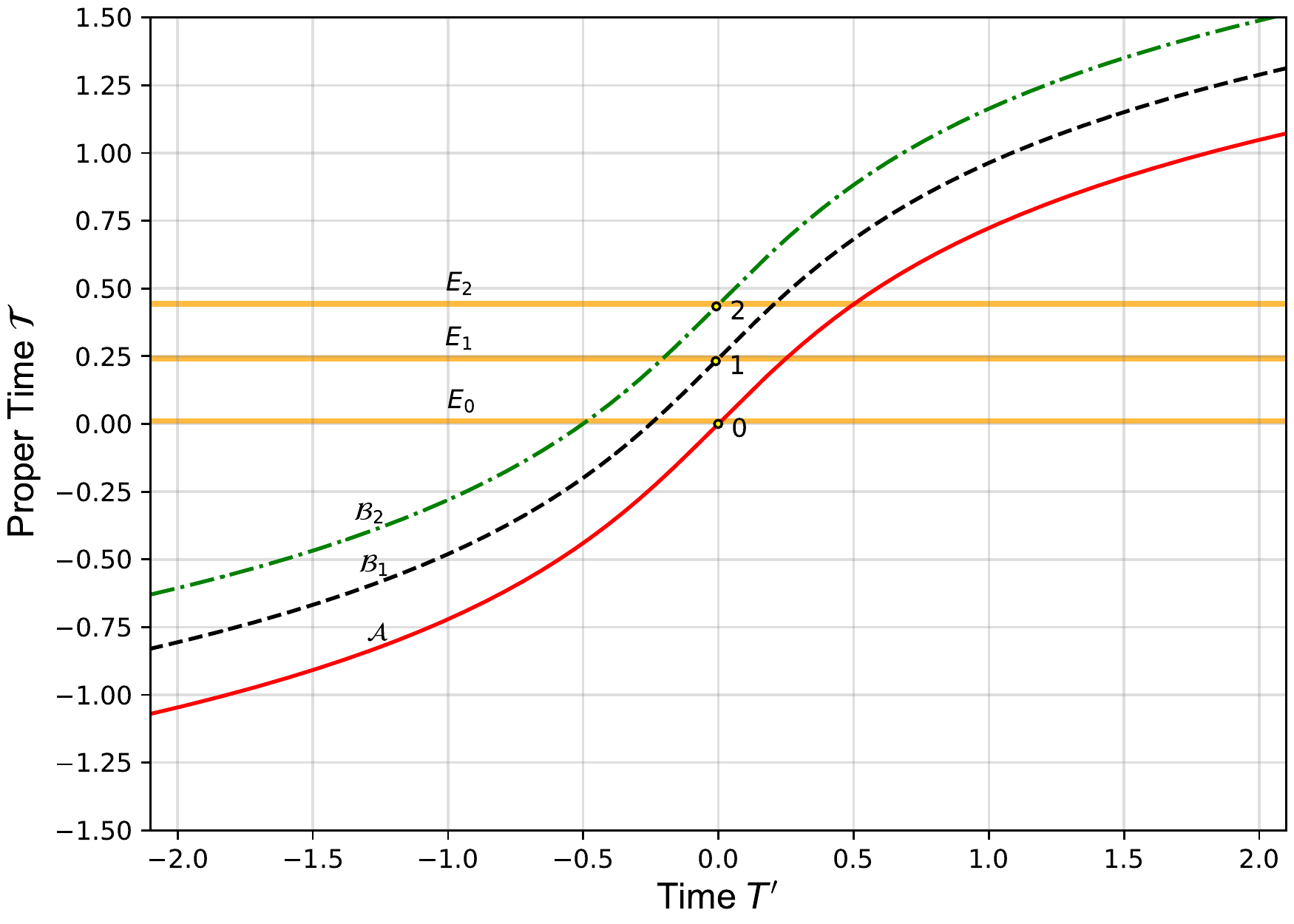}
    \caption{\scriptsize{Proper time $\tau$ vs Time $T^\prime$ measured by three inertial observers $\mathcal{A}~(u = 0.0)$, $\mathcal{B}_1~(u=0.45)$ and $\mathcal{B}_2~(u=0.71)$.}}
    \label{fig:3b}
\end{subfigure}
\caption{Proper time and Time dilation}
\label{fig:3}
\end{figure}

In Figure \ref{fig:3a} we plot the numerically deduced velocity $V^\prime$ of $\mathcal{R}$ measured by $\mathcal{B}$ with time $T^\prime$ recorded in the clock carried by the inertial observer $\mathcal{B}$ (again for $u_B=0.45$). The velocity curve is hyperbolic tangent in nature and we seek a parameter $\phi$ (``rapidity'') which varies linearly all along the Rindler worldline \cite{Rindler-1977, Rindler-2006}
\begin{equation}
    \phi = \tanh^{-1}\frac{V^\prime}{c}
\end{equation}
The Rindler observer carries an ideal clock that is not affected by its proper acceleration \cite{Rindler-2006,Sexl-2001}. In the frame of reference of $\mathcal{R}$, all the events $E$ are happening at the same place and the clock carried by $\mathcal{R}$ therefore measures proper time $\tau$ of these events \cite{Beiser-2003, Hassani-2017}. The relation between rapidity $\phi$ and proper time $\tau$ is \cite{Rindler-2006}
\begin{equation}\label{eq15}
    \tau = \frac{c}{a}\phi =  \frac{c}{a} \tanh^{-1}\left(\frac{V^\prime}{c}\right)
\end{equation}
We use the definition of rapidity to study the time dilation and so from Eq. \ref{eq15}, we also plot the proper time $\tau$ with respect to $T^\prime$ along with $\Delta \tau/\Delta T^\prime$ in Figure \ref{fig:3a}. Since $\Delta T^\prime \ge \Delta \tau$ at every time slice $T^\prime$, the clock of $\mathcal{R}$ appears to be moving slow or the time appears to be dilated for $\mathcal{B}$. We also find the time dilation factor $\Delta T^\prime/\Delta \tau$ depends on the velocity of $\mathcal{R}$ and verify that it equals the Lorentz Factor $\gamma_{\mathcal{R}}=(1-\beta_{\mathcal{R}}^2)^{-1/2}$ with $\beta_{\mathcal{R}}=V^\prime/c$. As shown in Fig. \ref{fig:3a}, at an arbitrary time slice $T^\prime=-0.63$ we find $\Delta T^\prime/\Delta \tau=1/0.63=1.59$.  This equals $\gamma_R=1.59$ and we can find that $V^\prime=\sqrt{1-\gamma^{-2}}=\pm 0.78$ which matches with the value of the speed in Fig \ref{fig:3a}. The curve for $V^\prime$ is independent of $u_B$ which is in accordance with Fig \ref{fig:2b}. Since all inertial observers $\mathcal{B}$ are equivalent, a measurement of time dilation factor $\Delta T^\prime/\Delta \tau$ for $\mathcal{R}$ should not differ for all these observers. Though $\Delta T^\prime/\Delta \tau$ is plotted for $u_B = 0.45$ in Fig \ref{fig:3a} it is independent of $u_B$.
Together with the code block used in Fig \ref{fig:2}, the following code snippet is used to obtain Fig \ref{fig:3a}.
\begin{lstlisting}[language = Python]
# Code for Fig 3a
# The variable t_slice_index can be modified between 0 and 2000 to obtain the index of an arbitrary time slice in B's frame.
t_slice_index = 970
tRI_slice = tRI[t_slice_index] #Value of time T' at the given index 
tau_slice = np.round(tau[t_slice_index],2) #Proper time at time T'
v_slice = np.round(vel[t_slice_index],2) #Velocity V' at time T'
t_dilation_slice = np.round(t_dilation[t_slice_index],2) #Time dilation factor at time T'
\end{lstlisting}
The proper time $\tau$ measured by the $\mathcal{R}$ when compared against $T^\prime$ measured by the $\mathcal{B}$ gets displaced depending on the speed $u_\mathcal{B}$. In Fig \ref{fig:3b} we compare $\tau$ with $T^\prime$ as measured by the three different inertial observers $\mathcal{A},~\mathcal{B}_1$ and $\mathcal{B}_2$ with speeds $0,~0.45$ and $0.71$ respectively. The observer $\mathcal{A}$ is only comoving with $\mathcal{R}$ at the point labelled $0$ where red solid curve has a slope of $45^\circ$. Similarly, at points $1$ and $2$ only, the clocks of observers $\mathcal{B}_1$ and $\mathcal{B}_2$ respectively run at the same rate as the ideal clock. The orange horizontal lines represent the fact that for a proper time $\tau$ the corresponding event $E$ is associated with a unique inertial observer $\mathcal{B}$. Fig \ref{fig:3b} is obtained by using the code block referred for Fig \ref{fig:2}.


\section{Conclusions}\label{sec5}
We use the point of view of an inertial reference frame and the measurements of successfully boosted inertial observers to elucidate properties of the accelerated observer's trajectory, viz, proper acceleration, instantaneously co-moving inertial frame and time dilation. We analyze this situation without resorting to the reference frame of the accelerated observer, thus making it easier for students to understand. While performing this classroom activity students find that for each event along the worldline of the Rindler observer we need a corresponding inertial observer that is instantaneously at rest. In other words, \textit{for every inertial observer B having appropriate speed u there exists a unique event E at which the observer is co-moving with Rindler observer $\mathcal{R}$}.  They also deduce that \textit{this event E is simultaneous to the event O and the measured distance of $\mathcal{B}$ to $\mathcal{R}$ is a minimum}. The instantaneous `proper' acceleration of $\mathcal{R}$ measured by such an observer is equal to the proper acceleration experienced by the $\mathcal{R}$. The multitude of inertial observers measure identical distance of closest approach and proper acceleration  as we explicitly show in Fig 2b. {\it The Rindler observer always has a proper acceleration $a=c^2/l_0$ where $l_0$ is the space-time interval of any event on the Rindler trajectory.}

\end{document}